\documentclass[12pt,a4paper]{article}
\usepackage{amsmath}
\usepackage{amssymb}
\usepackage{cite}

\renewcommand{\thesection}
  {\arabic{section}.\hspace{-.5em}}

\makeatletter
\renewcommand\section{
  \@startsection{section}{3}{\z@}%
  {-3.25ex\@plus -1ex \@minus -.2ex}%
  {1.5ex \@plus .2ex}%
  {\normalfont\normalsize\bfseries\mathversion{bold}}}
\renewcommand\subsection{
  \@startsection{subsection}{3}{\z@}%
  {-3.25ex\@plus -1ex \@minus -.2ex}%
  {1.5ex \@plus .2ex}%
  {\normalfont\normalsize\bfseries\mathversion{bold}}}
\makeatother

\makeatletter \@addtoreset{equation}{section} \makeatother
\renewcommand{\theequation}{\arabic{section}.\arabic{equation}}

\makeatletter
\renewcommand{\appendix}{
\renewcommand{\thesection}{\Alph{section}.\hspace{-.5em}}
\@addtoreset{equation}{section}
\renewcommand{\theequation}{\Alph{section}.\arabic{equation}}
\setcounter{section}{0}}
\makeatother

\let\oldthebibliography\thebibliography
\renewcommand\thebibliography[1]{
  \oldthebibliography{#1}\setlength{\itemsep}{0.4ex}}

\allowdisplaybreaks[1]

\newcommand{\atmark}{{\scriptsize
  \textcircled{\raisebox{-.15ex}{\small\it\hspace{-.12em}a}}}}

\newcommand{\Eqn}[1]{&\hspace{-0.5em}#1\hspace{-0.5em}&}
\newcommand{\nn}{\nonumber}
\renewcommand{\[}{\begin{eqnarray}}
\renewcommand{\]}{\end{eqnarray}}

\newcommand{\grp}[1]{\mathrm{#1}}
\newcommand{\bvec}[1]{\boldsymbol{#1}}
\newcommand{\varth}{\vartheta}

\newcommand{\bbC}{{\mathbb C}}
\newcommand{\bbR}{{\mathbb R}}
\newcommand{\bbZ}{{\mathbb Z}}

\newcommand{\vecal}{{\boldsymbol{\alpha}}}
\newcommand{\vece}{{\boldsymbol{\rm e}}}
\newcommand{\vecmu}{{\boldsymbol{\mu}}}
\newcommand{\vecL}{{\boldsymbol{\Lambda}}}
\newcommand{\vecv}{{\boldsymbol{v}}}
\newcommand{\vecw}{{\boldsymbol{w}}}
\newcommand{\veczero}{{\boldsymbol{0}}}

\newcommand{\WOEse}[7]{{}w[{}_{#1}^{}{}_{#3}^{}{}_{#4}^{#2}{}_{#5}^{}
                           {}_{#6}^{}{}_{#7}^{}]}
\newcommand{\WOEsi}[6]{{}w[{}_{#1}^{}{}_{#3}^{}{}_{#4}^{#2}{}_{#5}^{}
                           {}_{#6}^{}]}

\newcommand{\gen}{\alpha}

\begin{document}


\def\papertitlepage{\baselineskip 3.5ex \thispagestyle{empty}}
\def\preprinumber#1#2{\hfill
\begin{minipage}{1.22in}
#1 \par\noindent #2
\end{minipage}}

%
\papertitlepage
\setcounter{page}{0}
\preprinumber{arXiv:1706.04619}{}
\vskip 2ex
\vfill
\begin{center}
{\large\bf\mathversion{bold}
$E_n$ Jacobi forms and Seiberg--Witten curves
}
\end{center}
\vfill
\baselineskip=3.5ex
\begin{center}
Kazuhiro Sakai\\

{\small
\vskip 6ex
{\it Institute of Physics, Meiji Gakuin University,
Yokohama 244-8539, Japan}\\
\vskip 1ex
{\tt kzhrsakai\atmark gmail.com}

}
\end{center}
\vfill
\baselineskip=3.5ex
\begin{center} {\bf Abstract} \end{center}

We discuss Jacobi forms that are invariant under
the action of the Weyl group of type $E_n\ (n=6,7,8)$.
For $n=6,7$ we explicitly construct a full set of generators
of the algebra of $E_n$ weak Jacobi forms.
We first construct $n+1$ independent $E_n$ Jacobi forms
in terms of Jacobi theta functions and modular forms.
By using them we obtain Seiberg--Witten curves
of type $\tilde{E}_6$ and $\tilde{E}_7$ for the E-string theory.
The coefficients of each curve are $E_n$ weak Jacobi forms
of particular weights and indices specified by the root system,
realizing the generators whose existence 
was shown some time ago by Wirthm\"uller.

\vfill
\noindent
June 2017


\setcounter{page}{0}
\newpage
\renewcommand{\thefootnote}{\arabic{footnote}}
\setcounter{footnote}{0}
\setcounter{section}{0}
\baselineskip = 3.5ex
\pagestyle{plain}
%

\section{Introduction and summary}

The theory of Jacobi forms was first systematically studied
by Eichler and Zagier \cite{EichlerZagier}.
A Jacobi form is a holomorphic
function of complex variables $\tau$ and $\vecmu$
which has modular properties in $\tau$ and quasi-periodicity in
$\vecmu$. Jacobi forms invariant under the action of the Weyl group
$W(R)$ of a root system $R$ was investigated by Wirthm\"uller
\cite{Wirthmuller}.
Such Jacobi forms, which we call $W(R)$-invariant Jacobi forms
or just $R$ Jacobi forms, appear in various contexts
in mathematics and physics.

In \cite{Wirthmuller}
an inductive construction of the $W(R)$-invariant Jacobi forms
(except for $R=E_8$) was also presented.
The construction is, however, rather abstract
for $R=E_6,E_7$. On the other hand,
$W(E_8)$-invariant Jacobi forms were explicitly constructed
in the study of the E-string theory
\cite{Minahan:1998vr,Eguchi:2002fc,Sakai:2011xg}.
In \cite{Eguchi:2002fc} nine independent $E_8$ Jacobi forms
were first constructed in the course of deriving
the Seiberg--Witten curve for the E-string theory.
The construction was further refined in \cite{Sakai:2011xg}
in terms of concisely expressed $E_8$ holomorphic Jacobi forms.

In this paper we explicitly construct
a full set of generators of the algebra of 
$W(E_n)$-invariant weak Jacobi forms $(n=7,6)$.
We first construct $n+1$ independent $E_n$ holomorphic Jacobi forms.
Most of them are actually obtained
by mere reduction of $E_{n+1}$ Jacobi forms and thus
we have only to construct two new Jacobi forms
in each $E_n$ case. All these $E_n$ Jacobi forms are explicitly
expressed in terms of Jacobi theta functions and modular forms.

Using these Jacobi forms we next construct Seiberg--Witten curves
of type $\tilde{E}_7$ and $\tilde{E}_6$ for the E-string theory.
The original Seiberg--Witten curve for the E-string theory
is expressed in terms of $E_8$ Jacobi forms
\cite{Eguchi:2002fc,Sakai:2011xg}.
If we restrict the value of $\vecmu$
within the $E_n$ root space, the curve can be expressed
in terms of the above $n+1$ $E_n$ Jacobi forms.
We transform this curve into the form of the general
deformation of a singularity of type $\tilde{E}_n$.
The coefficients of this new Seiberg--Witten curve
are weak Jacobi forms of particular weights and indices
specified by the root system $E_n$.
They are identified as generators
of the algebra of $E_n$ weak Jacobi forms
over the algebra of modular forms.
The existence of such generators was shown
by Wirthm\"uller \cite{Wirthmuller}.

The main theorem of \cite{Wirthmuller} does not cover
the case of $R=E_8$. Very little has been known
about generators of the algebra of $E_8$ Jacobi forms
over the algebra of modular forms.
We briefly discuss this case
and make a conjecture on the overall picture of the algebra of
$E_8$ weak Jacobi forms.

The paper is organized as follows.
In section~2 we present the definition of $W(R)$-invariant Jacobi forms
and construct $n+1$ independent $E_n$ holomorphic Jacobi forms.
In section~3 we construct Seiberg--Witten curves
of type $\tilde{E}_7$ and $\tilde{E}_6$ for the E-string theory
and present a full set of generators of
the algebra of $E_n$ weak Jacobi forms for $n=7,6$.
We also discuss the case of $E_8$.
There are three appendices,
where Seiberg--Witten curves of type $\tilde{E}_n$
at $\tau=i\infty$,
our choice of simple roots and fundamental weights,
and definitions of special functions
are respectively presented.

\section{Construction of holomorphic Jacobi forms}

\subsection{Definitions and generalities}

Let $L_R$ be the root lattice of a root system $R$,
and $L_R^*$ the dual lattice of $L_R$.
Let $\varphi_{k,m}(\tau,\vecmu)$ denote
a $W(R)$-invariant Jacobi form of weight $k$ and index $m$
($k\in\bbZ,\ m\in\bbZ_{>0}$).
It is a holomorphic function of $\tau$ and $\vecmu$
(${\rm Im}\,\tau>0,\ \vecmu\in\bbC^n$) satisfying
the following properties \cite{EichlerZagier, Wirthmuller}:
\renewcommand{\theenumi}{\roman{enumi}}
\renewcommand{\labelenumi}{\theenumi)}
\begin{enumerate}
\item Weyl invariance:
\[\label{Weylinv}
\varphi_{k,m}(\tau,w(\vecmu)) = \varphi_{k,m}(\tau,\vecmu),\qquad
w\in W(R).
\]

\item Quasi-periodicity:
\[
\varphi_{k,m}(\tau,\vecmu+\tau\bvec{\alpha}+\bvec{\beta})
=e^{-m \pi i (\tau\bvec{\alpha}^2+2\vecmu\cdot\bvec{\alpha})}
\varphi_{k,m}(\tau,\vecmu),\qquad
\bvec{\alpha},\bvec{\beta}\in L_R.
\]

\item Modular properties:
\begin{align}\label{Modularprop}
&
\varphi_{k,m}\left(
\frac{a\tau+b}{c\tau+d}\,,\frac{\vecmu}{c\tau+d}\right)
=(c\tau+d)^k\exp\left(m\pi i\frac{c}{c\tau+d}\,\vecmu^2\right)
\varphi_{k,m}(\tau,\vecmu),\\[1ex]
&
\Bigl(\begin{array}{cc}a&b\\ c&d\end{array}\Bigr)
\in \grp{SL}(2,\bbZ).\nn
\end{align}

\item $\varphi_{k,m}(\tau,\vecmu)$ admits a Fourier expansion as
\[\label{Fourierform}
\varphi_{k,m}(\tau,\vecmu)
=\sum_{n=0}^\infty
 \sum_{\bvec{w}\in L_R^*}
 c(n,\bvec{w})e^{2\pi i(n\tau+\bvec{w}\cdot\vecmu)}.
\]

\end{enumerate}
To be precise, $\varphi_{k,m}(\tau,\vecmu)$ defined
as above is called a weak Jacobi form.
If $\varphi_{k,m}(\tau,\vecmu)$ further satisfies the condition
that the coefficients $c(n,\bvec{w})$
of the Fourier expansion (\ref{Fourierform})
vanish unless $\bvec{w}^2\le 2mn$,
it is called a holomorphic Jacobi form.
If $\varphi_{k,m}(\tau,\vecmu)$ further satisfies the stronger
condition that the coefficients
$c(n,\bvec{w})$ vanish unless $\bvec{w}^2< 2mn$,
it is called a Jacobi cusp form.
In this paper a Jacobi form means a weak Jacobi form
unless otherwise specified.

The condition (\ref{Weylinv}) and
the form of the Fourier expansion (\ref{Fourierform}) imply that
$W(R)$-invariant Jacobi forms are closely related to
characters of Weyl orbits of the affine $R$ Lie algebra.
In our convention the index coincides with the level
of the affine Lie algebra.
In fact, 
we observe that any $W(R)$-invariant Jacobi form of index $m$
can be written as a linear combination of characters of
affine Weyl orbits of level-$m$ weights, and vice versa.
From this one can expect that
the number of generators of Jacobi forms of index $m$
coincides with the number of fundamental representations
at level $m$.\footnote{Here, ``generators'' 
do not mean those for the algebra of $W(R)$-invariant Jacobi forms
over the ring of modular forms $\bbC[E_4,E_6]$.
Instead, we consider here a bigger space where
we allow meromorphic modular forms as coefficients.}
Figure \ref{Fig:Dynkin} shows the levels
of fundamental representations of the affine $E_n$ algebra.
\begin{figure}[t]
\begin{center}\vspace{-.1ex}
\unitlength=2.4pt
\begin{picture}(60,30)
\put(-15,9){$\hat{E}_8$:}
\put( 0,10){\circle{2}}
\put( 0,10){${\!}^2$}
\put( 0, 6){${\!\!\!\!\>}_{(1)}$}
\put( 1,10){\line(1,0){8}}
\put(10,10){\circle{2}}
\put(10,10){${\!}^4$}
\put(10, 6){${\!\!\!\!\>}_{(3)}$}
\put(11,10){\line(1,0){8}}
\put(20,10){\circle{2}}
\put(20,10){${\,}^6$}
\put(20, 6){${\!\!\!\!\>}_{(4)}$}
\put(21,10){\line(1,0){8}}
\put(30,10){\circle{2}}
\put(30,10){${\!}^5$}
\put(30, 6){${\!\!\!\!\>}_{(5)}$}
\put(31,10){\line(1,0){8}}
\put(40,10){\circle{2}}
\put(40,10){${\!}^4$}
\put(40, 6){${\!\!\!\!\>}_{(6)}$}
\put(41,10){\line(1,0){8}}
\put(50,10){\circle{2}}
\put(50,10){${\!}^3$}
\put(50, 6){${\!\!\!\!\>}_{(7)}$}
\put(51,10){\line(1,0){8}}
\put(60,10){\circle{2}}
\put(60,10){${\!}^2$}
\put(60, 6){${\!\!\!\!\>}_{(8)}$}
\put(61,10){\line(1,0){8}}
\put(70,10){\circle{2}}
\put(70,10){${\!}^1$}
\put(70, 6){${\!\!\!\!\>}_{(0)}$}
\put(20,11){\line(0,1){8}}
\put(20,20){\circle{2}}
\put(20,20){${\!}^3$}
\put(20,20){${\hspace{0.4em}}_{(2)}$}
\end{picture}

\begin{picture}(60,30)
\put(-15,9){$\hat{E}_7$:}
\put( 0,10){\circle{2}}
\put( 0,10){${\!}^1$}
\put( 0, 6){${\!\!\!\!\>}_{(0)}$}
\put( 1,10){\line(1,0){8}}
\put(10,10){\circle{2}}
\put(10,10){${\!}^2$}
\put(10, 6){${\!\!\!\!\>}_{(1)}$}
\put(11,10){\line(1,0){8}}
\put(20,10){\circle{2}}
\put(20,10){${\!}^3$}
\put(20, 6){${\!\!\!\!\>}_{(3)}$}
\put(21,10){\line(1,0){8}}
\put(30,10){\circle{2}}
\put(30,10){${\,}^4$}
\put(30, 6){${\!\!\!\!\>}_{(4)}$}
\put(31,10){\line(1,0){8}}
\put(40,10){\circle{2}}
\put(40,10){${\!}^3$}
\put(40, 6){${\!\!\!\!\>}_{(5)}$}
\put(41,10){\line(1,0){8}}
\put(50,10){\circle{2}}
\put(50,10){${\!}^2$}
\put(50, 6){${\!\!\!\!\>}_{(6)}$}
\put(51,10){\line(1,0){8}}
\put(60,10){\circle{2}}
\put(60,10){${\!}^1$}
\put(60, 6){${\!\!\!\!\>}_{(7)}$}
\put(30,11){\line(0,1){8}}
\put(30,20){\circle{2}}
\put(30,20){${\!}^2$}
\put(30,20){${\hspace{0.4em}}_{(2)}$}
\end{picture}

\begin{picture}(60,40)
\put(-15,19){$\hat{E}_6$:}
\put(10,10){\circle{2}}
\put(10,10){${\!}^1$}
\put(10, 6){${\!\!\!\!\>}_{(1)}$}
\put(11,10){\line(1,0){8}}
\put(20,10){\circle{2}}
\put(20,10){${\!}^2$}
\put(20, 6){${\!\!\!\!\>}_{(3)}$}
\put(21,10){\line(1,0){8}}
\put(30,10){\circle{2}}
\put(30,10){${\!\!\!}^3$}
\put(30, 6){${\!\!\!\!\>}_{(4)}$}
\put(31,10){\line(1,0){8}}
\put(40,10){\circle{2}}
\put(40,10){${\!}^2$}
\put(40, 6){${\!\!\!\!\>}_{(5)}$}
\put(41,10){\line(1,0){8}}
\put(50,10){\circle{2}}
\put(50,10){${\!}^1$}
\put(50, 6){${\!\!\!\!\>}_{(6)}$}
\put(30,11){\line(0,1){8}}
\put(30,20){\circle{2}}
\put(30,20){${\hspace{-.7em}}_{2}$}
\put(30,20){${\hspace{0.4em}}_{(2)}$}
\put(30,21){\line(0,1){8}}
\put(30,30){\circle{2}}
\put(30,30){${\hspace{-.7em}}_{1}$}
\put(30,30){${\hspace{0.4em}}_{(0)}$}
\end{picture}
\caption{Dynkin diagram for affine $E_n$:
numbers attached to nodes denote the levels of fundamental weights
and the numbers in parentheses show their labels.
\label{Fig:Dynkin}}
\end{center}
\end{figure}
From this we see that
generators of $R$ Jacobi forms are of the indices
\[\label{indlist}
1,2,2,3,3,4,4,5,6\ &\mbox{for}& E_8,\nn\\
1,1,2,2,2,3,3,4\ &\mbox{for}& E_7,\nn\\
1,1,1,2,2,2,3\ &\mbox{for}& E_6.
\]
Multiple occurrence of the same index means that
there are several independent generators of the index.
In what follows we will explicitly construct $E_n$ Jacobi forms
of these indices.

\subsection{$E_8$ case}

Nine independent $W(E_8)$-invariant holomorphic Jacobi forms
were constructed in \cite{Sakai:2011xg}.
The summary of the results is shown below.

Let us first introduce the following functions
\[
e_1(\tau)\Eqn{:=}
 \tfrac{1}{12}\left(\varth_3(\tau)^4+\varth_4(\tau)^4\right),\nn\\
e_2(\tau)\Eqn{:=}
 \tfrac{1}{12}\left(\varth_2(\tau)^4-\varth_4(\tau)^4\right),\nn\\
e_3(\tau)\Eqn{:=}
 \tfrac{1}{12}\left(-\varth_2(\tau)^4-\varth_3(\tau)^4\right),
\]
and
\[\label{h0def}
h_0(\tau)
  :=\varth_3(2\tau)\varth_3(6\tau)+\varth_2(2\tau)\varth_2(6\tau).
\]
The simplest $E_8$ Jacobi form is
the theta function of the root lattice $L_{E_8}$:
\begin{align}
\Theta_{E_8}(\tau,\vecmu)
:=&\sum_{\vecw\in L_{E_8}}
   \exp\left(\pi i\tau\vecw^2
   +2\pi i\vecmu\cdot\vecw\right)\\
=&\ \frac{1}{2}\sum_{k=1}^4\prod_{j=1}^8\varth_k(\mu_j,\tau).
\end{align}
Nine $W(E_8)$-invariant holomorphic Jacobi forms
can be constructed as follows:
\[\label{E8AB}
A_1(\tau,\vecmu)\Eqn{=}\Theta_{E_8}(\tau,\vecmu),\qquad
A_4(\tau,\vecmu)=A_1(\tau,2\vecmu),\nn\\
A_m(\tau,\vecmu)\Eqn{=}\tfrac{m^3}{m^3+1}\left(
  A_1(m\tau,m\vecmu)
  +\tfrac{1}{m^4}\mbox{$\sum_{k=0}^{m-1}$}
  A_1(\tfrac{\tau+k}{m},\vecmu)
 \right),\qquad m=2,3,5,\nn\\
B_2(\tau,\vecmu)\Eqn{=}\tfrac{32}{5}\left(
 e_1(\tau)A_1(2\tau,2\vecmu)
 +\tfrac{1}{2^4}e_3(\tau)A_1(\tfrac{\tau}{2},\vecmu)
 +\tfrac{1}{2^4}e_2(\tau)A_1(\tfrac{\tau+1}{2},\vecmu)\right),\nn\\
B_3(\tau,\vecmu)\Eqn{=}\tfrac{81}{80}\left(
 h_0(\tau)^2A_1(3\tau,3\vecmu)
  -\tfrac{1}{3^5}\mbox{$\sum_{k=0}^{2}$}h_0(\tfrac{\tau+k}{3})^2
  A_1(\tfrac{\tau+k}{3},\vecmu)\right),\nn\\
B_4(\tau,\vecmu)\Eqn{=}\tfrac{16}{15}\left(
 \varth_4(2\tau)^4A_1(4\tau,4\vecmu)
 -\tfrac{1}{2^4}\varth_4(2\tau)^4
  A_1(\tau+\tfrac{1}{2},2\vecmu)\right.\nn\\
&&\hspace{2em}
 \left.
 -\tfrac{1}{2^2\cdot 4^4}\mbox{$\sum_{k=0}^{3}$}
  \varth_2(\tfrac{\tau+k}{2})^4
  A_1(\tfrac{\tau+k}{4},\vecmu)\right),\nn\\
B_6(\tau,\vecmu)\Eqn{=}\tfrac{9}{10}\left(
  h_0(\tau)^2A_1(6\tau,6\vecmu)
 +\tfrac{1}{2^4}\mbox{$\sum_{k=0}^{1}$}
  h_0(\tau+k)^2A_1(\tfrac{3\tau+3k}{2},3\vecmu)\right.\nn\\
&&\hspace{2em}\left.
 -\tfrac{1}{3\cdot 3^4}\mbox{$\sum_{k=0}^{2}$}
  h_0(\tfrac{\tau+k}{3})^2A_1(\tfrac{2\tau+2k}{3},2\vecmu)\right.\nn\\
&&\hspace{2em}\left.
 -\tfrac{1}{3\cdot 6^4}\mbox{$\sum_{k=0}^{5}$}
  h_0(\tfrac{\tau+k}{3})^2A_1(\tfrac{\tau+k}{6},\vecmu)\right).
\]
$A_m,B_m$ are of weight $4,6$ and index $m$ respectively.
If we set $\vecmu=\veczero$,
these Jacobi forms reduce to ordinary modular forms.
The normalization of these Jacobi forms is chosen
so that they reduce to the Eisenstein series
\[
A_m(\tau,\veczero)=E_4(\tau),\qquad
B_m(\tau,\veczero)=E_6(\tau).
\]
For the sake of clarity
the above $A_m,B_m$ are sometimes expressed as $A_m^{E_8},B_m^{E_8}$.

\subsection{$E_7$ case}

$W(E_7)$-invariant Jacobi forms can be obtained
by reduction of $W(E_8)$-invariant ones.
This is done by merely restricting $\vecmu$ within
the $E_7$ root space orthogonal to
the fundamental weight $\vecL_8^{E_8}$.
More specifically, such $\vecmu$ is parametrized as
\[
\vecmu=\vecmu^{(7)}:=(\mu_1,\mu_2,\mu_3,\mu_4,\mu_5,\mu_6,\mu,-\mu).
\]
See Appendix~B for our convention.
In what follows in this subsection $\vecmu$ is
always constrained as above.

By reducing the $E_8$ Jacobi forms given in (\ref{E8AB})
one immediately obtains
\[\label{E8E7reduction}
A_m^{E_7}(\tau,\vecmu)\Eqn{:=}A_m^{E_8}\left(\tau,\vecmu^{(7)}\right)
 \qquad (m=1,2,3,4,5),\nn\\
B_m^{E_7}(\tau,\vecmu)\Eqn{:=}B_m^{E_8}\left(\tau,\vecmu^{(7)}\right)
 \qquad (m=2,3,4,6).
\]
These Jacobi forms cover most of the desired $E_7$ Jacobi forms
whose indices are listed in (\ref{indlist}), but not all of them.
We need to construct in addition
at least two new Jacobi forms which are of index one and index two
respectively.

Let us start our study with $E_7$ Jacobi forms of index one.
There are two independent $E_7$ Jacobi forms.
They can be expressed as some modular-invariant
linear combinations of
two level-one affine Weyl orbit characters.
At level one, affine Weyl orbit characters
are simply given by the theta functions
\[
\Theta_{E_7}(\tau,\vecmu)
\Eqn{:=}
\sum_{\vecw\in L_{E_7}}
   \exp\left(\pi i\tau\vecw^2
   +2\pi i\vecmu\cdot\vecw\right),\nn\\
\Theta_{E_7}^{[7]}(\tau,\vecmu)
\Eqn{:=}
\sum_{\vecw\in L_{E_7}+\vecL_7}
   \exp\left(\pi i\tau\vecw^2
   +2\pi i\vecmu\cdot\vecw\right).
\]
Here, $\vecL_7=\vecL_7^{E_7}$ is a fundamental weight
of $E_7$. (See Appendix B.)
In terms of Jacobi theta functions they are expressed as
\[
\Theta_{E_7}\Eqn{=}
 \frac{1}{2}\varth_2(2\mu,2\tau)
 \sum_{k=1}^2\prod_{j=1}^6\varth_k(\mu_j,\tau)
+\frac{1}{2}\varth_3(2\mu,2\tau)
 \sum_{k=3}^4\prod_{j=1}^6\varth_k(\mu_j,\tau),\nn\\
\Theta_{E_7}^{[7]}\Eqn{=}
 \frac{1}{2}\varth_3(2\mu,2\tau)
 \sum_{k=1}^2(-1)^k\prod_{j=1}^6\varth_k(\mu_j,\tau)
-\frac{1}{2}\varth_2(2\mu,2\tau)
 \sum_{k=3}^4(-1)^k\prod_{j=1}^6\varth_k(\mu_j,\tau).\nn\\
\]
Note that Fourier expansions of these theta functions are
\[
\Theta_{E_7}\Eqn{=}
 1+\WOEse{1}{0}{0}{0}{0}{0}{0}q
  +\WOEse{0}{0}{0}{0}{0}{1}{0}q^2
  +\left(\WOEse{0}{0}{1}{0}{0}{0}{0}
  +\WOEse{0}{0}{0}{0}{0}{0}{2}\right)
   q^3+{\cal O}(q^4),\nn\\
\Theta_{E_7}^{[7]}\Eqn{=}
  \WOEse{0}{0}{0}{0}{0}{0}{1}q^{3/4}
 +\WOEse{0}{1}{0}{0}{0}{0}{0}q^{7/4}
 +\WOEse{1}{0}{0}{0}{0}{0}{1}q^{11/4}
 +\WOEse{0}{0}{0}{0}{1}{0}{0}q^{15/4}
 +{\cal O}(q^{19/4}),\nn\\
\]
where $q:=e^{2\pi i\tau}$.
The coefficients are expressed in terms of
characters of Weyl orbits of finite $E_7$.
They are defined by
\[
\WOEse{n_1}{n_2}{n_3}{n_4}{n_5}{n_6}{n_7}(\vecmu):=
 \sum_{\vecv\in{\cal O}(\sum_{j=1}^7n_j\vecL_j)}
 e^{2\pi i\vecv\cdot\vecmu}.
\]
Here, ${\cal O}(\vecL)$ denotes
the Weyl orbit of weight $\vecL$. $\vecL_j\ (j=1,\ldots,7)$
are the fundamental weights of $E_7$.

The above theta functions transform nontrivially under modular
transformations.
The modular properties of the theta functions are as follows:
\[
\Theta_{E_7}\left(\tau+1,\vecmu\right)
 \Eqn{=}\Theta_{E_7}\left(\tau,\vecmu\right),\nn\\
\Theta_{E_7}^{[7]}\left(\tau+1,\vecmu\right)
 \Eqn{=}-i\Theta_{E_7}^{[7]}\left(\tau,\vecmu\right),\\[1ex]
\left(\begin{array}{c}
\Theta_{E_7}      \left(-\frac{1}{\tau},\frac{\vecmu}{\tau}\right)
\\[1ex]
\Theta_{E_7}^{[7]}\left(-\frac{1}{\tau},\frac{\vecmu}{\tau}\right)
\end{array}\right)
\Eqn{=}
e^{-\frac{7\pi i}{4}}
\tau^{\frac{7}{2}}
e^{\frac{\pi i}{\tau}\vecmu^2}
\frac{1}{\sqrt{2}}
\left(
\begin{array}{cc}
1&1\\
1&-1
\end{array}
\right)
\left(\begin{array}{c}
\Theta_{E_7}      \left(\tau,\vecmu\right)\\[1ex]
\Theta_{E_7}^{[7]}\left(\tau,\vecmu\right)
\end{array}\right).\quad
\]
To construct modular-invariant linear
combinations of $\Theta_{E_7}$ and $\Theta_{E_7}^{[7]}$,
let us first look into the case of $A_1^{E_7}$.
One can easily derive that $A_1^{E_7}$ is expressed as
\[\label{A1E7}
A_1^{E_7}(\tau,\vecmu)\Eqn{=}
 \varth_3(2\tau)\Theta_{E_7}(\tau,\vecmu)
+\varth_2(2\tau)\Theta_{E_7}^{[7]}(\tau,\vecmu).
\]
The coefficient functions can be interpreted as
$\varth_3(2\tau)=\Theta_{A_1}(\tau,0)$,
$\varth_2(2\tau)=\Theta_{A_1}^{[1]}(\tau,0)$
and transform as
\[
\varth_3(2(\tau+1))\Eqn{=}\varth_3(2\tau),\qquad
\varth_2(2(\tau+1))=i\varth_2(2\tau),\nn\\[1ex]
\left(\begin{array}{c}
\varth_3\left(-\frac{2}{\tau}\right)\\[1ex]
\varth_2\left(-\frac{2}{\tau}\right)
\end{array}\right)
\Eqn{=}
e^{-\frac{\pi i}{4}}
\tau^{\frac{1}{2}}
\frac{1}{\sqrt{2}}
\left(
\begin{array}{cc}
1&1\\
1&-1
\end{array}
\right)
\left(\begin{array}{c}
\varth_3\left(2\tau\right)\\[1ex]
\varth_2\left(2\tau\right)
\end{array}\right).\quad
\]
One can easily check that
(\ref{A1E7}) is indeed a modular-invariant combination,
i.e.~it transforms as in (\ref{Modularprop}).

It is natural to expect that the other modular-invariant
linear combination can also be constructed by using polynomials of
$\varth_3(2\tau),\varth_2(2\tau)$ as coefficient functions.
One of the simplest candidates for this Jacobi form would be the one
which reduces to $E_6(\tau)$ when we set $\vecmu=\veczero$.
In order for the Jacobi form to be of weight 6,
the coefficient functions have to be homogeneous quintics
in $\varth_3(2\tau),\varth_2(2\tau)$.
And furthermore,
in order to be invariant under the transformation $\tau\to\tau+1$,
the Jacobi form has to take the form
\[
 \left(c_1\varth_3(2\tau)^4+c_2\varth_2(2\tau)^4\right)\varth_3(2\tau)
 \Theta_{E_7}
+\left(c_3\varth_3(2\tau)^4+c_4\varth_2(2\tau)^4\right)\varth_2(2\tau)
 \Theta_{E_7}^{[7]}.
\]
The requirement that it reduces to $E_6$ when $\vecmu=\veczero$
immediately determines the unknown coefficients $c_j$.
In this way, we find the combination
\[
C_1^{E_7}(\tau,\vecmu)\Eqn{:=}
 \left(\varth_3(2\tau)^4-5\varth_2(2\tau)^4\right)
 \varth_3(2\tau)\Theta_{E_7}(\tau,\vecmu)\nn\\
&&
+\left(\varth_2(2\tau)^4-5\varth_3(2\tau)^4\right)
 \varth_2(2\tau)\Theta_{E_7}^{[7]}(\tau,\vecmu).
\]
One can check that $C_1^{E_7}$ is indeed
an $E_7$ holomorphic Jacobi form of index one.
It is clear that $A_1^{E_7}$ and $C_1^{E_7}$ are independent.
By construction,
\[
C_1^{E_7}(\tau,\veczero)=E_6(\tau).
\]

Let us now move on to the construction of a new Jacobi form of index two.
This is actually easy.
Applying the Hecke transformation of order two to $C_1^{E_7}$,
one obtains
\[
C_2^{E_7}(\tau,\vecmu)
 \Eqn{:=}\frac{32}{33}\left(
  C_1^{E_7}(2\tau,2\vecmu)
  +\frac{1}{64}\sum_{k=0}^1
  C_1^{E_7}\left(\frac{\tau+k}{2},\vecmu\right)
 \right).
\]
The normalization is chosen so that
\[
C_2^{E_7}(\tau,\veczero)=E_6(\tau).
\]
One can check that $C_2^{E_7}$ is an independent Jacobi form,
i.e.~it is not expressed
as polynomials in $A_1^{E_7}, C_1^{E_7}, A_2^{E_7}, B_2^{E_7}$.

One can also check that
$A_3^{E_7},B_3^{E_7}$ are independent in the same sense.
On the other hand, it turns out that $A_4^{E_7}$ is not independent.
It is expressed in terms of
$A_m^{E_7}$, $B_m^{E_7}$, $C_m^{E_7}\ (m\le 3)$
as
\[\label{E7A4rel}
A_4
\Eqn{=}\frac{1}{13824E_4^2\Delta}
\left(
-448E_4^4A_1A_3+448E_4^2E_6C_1A_3
-1280E_4^2E_6A_1B_3+1280E_4^3C_1B_3\right.\nn\\
&&\hspace{1.5em}
+216E_4^4A_2^2
-1440E_4^3A_1^2A_2+720E_4^2E_6A_2B_2+288E_4^2C_1^2A_2\nn\\
&&\hspace{1.5em}
+(1275E_4^3-675E_6^2)B_2^2
+(-990E_4^3+990E_6^2)B_2C_2+360E_4E_6A_1^2B_2\nn\\
&&\hspace{1.5em}
-2640E_4^2A_1C_1B_2+360E_6C_1^2B_2
+(363E_4^3-363E_6^2)C_2^2-264E_4E_6A_1^2C_2\nn\\
&&\hspace{1.5em}
+528E_4^2A_1C_1C_2-264E_6C_1^2C_2
+1680E_4^2A_1^4-96E_4A_1^2C_1^2-48C_1^4\left.\right).
\]
Here, we have omitted superscript $E_7$ from the Jacobi forms and
introduced
\[
\Delta := \eta^{24}=\frac{1}{1728}\left(E_4^3-E_6^2\right).
\]

To summarize, we now have eight Jacobi forms
\[\label{E7ABC}
A_m^{E_7}\ \ (m=1,2,3),\quad
B_m^{E_7}\ \ (m=2,3,4),\quad
C_m^{E_7}\ \ (m=1,2),
\]
which are of weight $4,6,6$ and index $m$ respectively.
We checked that they are independent, holomorphic Jacobi forms.
Note that
\[
A_m^{E_7}(\tau,\veczero)=E_4(\tau),\qquad
B_m^{E_7}(\tau,\veczero)=C_m^{E_7}(\tau,\veczero)=E_6(\tau).
\]

As expected, $A_5^{E_7}, B_6^{E_7}$ are no longer independent
and are expressed as polynomials
in the eight Jacobi forms (\ref{E7ABC}).
While these relations are essential to obtain the results
in the next section, their concrete expressions are rather lengthy
and thus we do not present them here.
(In any case, these relations are immediately restored
from the results in the next section.)

\subsection{$E_6$ case}

As in the $E_7$ case, $W(E_6)$-invariant Jacobi forms
can be obtained by reduction of those for $E_7$ or $E_8$.
This is done by restricting $\vecmu$ within
the $E_6$ root space orthogonal to
both $\vecL_7^{E_8}$ and $\vecL_8^{E_8}$.
More specifically, such a vector $\vecmu$ is parametrized as
\[
\vecmu=\vecmu^{(6)}:=(\mu_1,\mu_2,\mu_3,\mu_4,\mu_5,\mu,\mu,-\mu).
\]
In what follows in this subsection $\vecmu$ is always constrained
as above.

By reducing the $E_7$ (or $E_8$) Jacobi forms
one immediately obtains
\begin{align}
\label{E7E6reduction}
&&A_m^{E_6}(\tau,\vecmu)&:=
  A_m^{E_7}\left(\tau,\vecmu^{(6)}\right)
 =A_m^{E_8}\left(\tau,\vecmu^{(6)}\right)&&(m=1,2,3),\nn\\
&&B_m^{E_6}(\tau,\vecmu)&:=
  B_m^{E_7}\left(\tau,\vecmu^{(6)}\right)
 =B_m^{E_8}\left(\tau,\vecmu^{(6)}\right)&&(m=2,3,4),\nn\\
&&C_m^{E_6}(\tau,\vecmu)&:=
  C_m^{E_7}\left(\tau,\vecmu^{(6)}\right)&&(m=1,2).
\end{align}
These Jacobi forms cover most of
the desired $E_6$ Jacobi forms
whose indices are listed in (\ref{indlist}), but 
as in the $E_7$ case, we need to construct
at least two new Jacobi forms
which are of index one and index two
respectively.

There are three affine $E_6$ Weyl orbit characters at level one:
$\Theta_{E_6}, \Theta_{E_6}^{[1]}$ and $\Theta_{E_6}^{[6]}$.
They are defined
by means of the root lattice $E_6$ and fundamental weights
$\vecL_1^{E_6}, \vecL_6^{E_6}$
in the same way as in the $E_7$ case.
(See Appendix~B for our convention.)
They are expressed in terms of Jacobi theta functions as
\[
\Theta_{E_6}(\tau,\vecmu)\Eqn{=}
 \frac{1}{2}\sum_{k=1}^4
  \varth_k(3\mu,3\tau)\prod_{j=1}^5\varth_k(\mu_j,\tau),\nn\\
\Theta_{E_6}^{[1]}(\tau,\vecmu)\Eqn{=}
 \frac{1}{2}\sum_{k=1}^4\sigma(k)q^{1/6}e^{2\pi i\mu}
  \varth_k(3\mu+\tau,3\tau)\prod_{j=1}^5\varth_k(\mu_j,\tau),\nn\\
\Theta_{E_6}^{[6]}(\tau,\vecmu)\Eqn{=}
 \frac{1}{2}\sum_{k=1}^4\sigma(k)q^{1/6}e^{-2\pi i\mu}
  \varth_k(3\mu-\tau,3\tau)\prod_{j=1}^5\varth_k(\mu_j,\tau),
\]
where $\sigma(1)=\sigma(4)=-1,\ \sigma(2)=\sigma(3)=1$.
These theta functions are expanded as
\[
\Theta_{E_6}\Eqn{=}
 1+\WOEsi{0}{1}{0}{0}{0}{0}q+\WOEsi{1}{0}{0}{0}{0}{1}q^2
  +\WOEsi{0}{0}{0}{1}{0}{0}q^3+{\cal O}(q^4),\nn\\
\Theta_{E_6}^{[1]}\Eqn{=}
  \WOEsi{1}{0}{0}{0}{0}{0}q^{2/3}
 +\WOEsi{0}{0}{0}{0}{1}{0}q^{5/3}
 +\left(\WOEsi{1}{1}{0}{0}{0}{0}
 +\WOEsi{0}{0}{0}{0}{0}{2}\right)q^{8/3}
 +{\cal O}(q^{11/3}),\nn\\
\Theta_{E_6}^{[6]}\Eqn{=}
  \WOEsi{0}{0}{0}{0}{0}{1}q^{2/3}
 +\WOEsi{0}{0}{1}{0}{0}{0}q^{5/3}
 +\left(\WOEsi{0}{1}{0}{0}{0}{1}
 +\WOEsi{2}{0}{0}{0}{0}{0}\right)q^{8/3}
 +{\cal O}(q^{11/3}).\quad
\]
The modular properties of these theta functions are as follows:
\[
\Theta_{E_6}\left(\tau+1,\vecmu\right)
 \Eqn{=}\Theta_{E_6}\left(\tau,\vecmu\right),\nn\\
\Theta_{E_6}^{[1]}\left(\tau+1,\vecmu\right)
 \Eqn{=}e^{4\pi i/3}\Theta_{E_6}^{[1]}\left(\tau,\vecmu\right),\nn\\
\Theta_{E_6}^{[6]}\left(\tau+1,\vecmu\right)
 \Eqn{=}e^{4\pi i/3}\Theta_{E_6}^{[6]}\left(\tau,\vecmu\right),\\[1ex]
\left(\begin{array}{c}
\Theta_{E_6}      \left(-\frac{1}{\tau},\frac{\vecmu}{\tau}\right)
\\[1ex]
\Theta_{E_6}^{[1]}\left(-\frac{1}{\tau},\frac{\vecmu}{\tau}\right)
\\[1ex]
\Theta_{E_6}^{[6]}\left(-\frac{1}{\tau},\frac{\vecmu}{\tau}\right)
\end{array}\right)
\Eqn{=}
i\tau^3e^{\frac{\pi i}{\tau}\vecmu^2}
\frac{1}{\sqrt{3}}
\left(
\begin{array}{ccc}
1&1&1\\
1&e^{4\pi i/3}&e^{2\pi i/3}\\
1&e^{2\pi i/3}&e^{4\pi i/3}
\end{array}
\right)
\left(\begin{array}{c}
\Theta_{E_6}      \left(\tau,\vecmu\right)\\[1ex]
\Theta_{E_6}^{[1]}\left(\tau,\vecmu\right)\\[1ex]
\Theta_{E_6}^{[6]}\left(\tau,\vecmu\right)
\end{array}\right).\qquad
\]

$A_1^{E_6}$ and $C_1^{E_6}$ are expressed
in terms of these theta functions as
\[
A_1^{E_6}(\tau,\vecmu)\Eqn{=}
 h_0\Theta_{E_6}(\tau,\vecmu)
+h_1
 \left(\Theta_{E_6}^{[1]}(\tau,\vecmu)
      +\Theta_{E_6}^{[6]}(\tau,\vecmu)\right),\\
C_1^{E_6}(\tau,\vecmu)\Eqn{=}
 \left(h_0^3-4h_1^3\right)\Theta_{E_6}(\tau,\vecmu)
-3h_0^2h_1
 \left(\Theta_{E_6}^{[1]}(\tau,\vecmu)
      +\Theta_{E_6}^{[6]}(\tau,\vecmu)\right),
\]
where $h_j=h_j(\tau)$. $h_0(\tau)$ was introduced
in (\ref{h0def}) and
\[
h_1(\tau)\Eqn{:=}
  3\frac{\eta(3\tau)^3}{\eta(\tau)}
  =\frac{1}{2}\left(h_0(\tau/3)-h_0(\tau)\right).
\]
They can be interpreted as
$h_0(\tau)=\Theta_{A_2}(\tau,\veczero)$,
$h_1(\tau)=\Theta_{A_2}^{[1]}(\tau,\veczero)
          =\Theta_{A_2}^{[2]}(\tau,\veczero)$.

By taking account of the above modular properties,
the other Jacobi form of index one is found as
\[
D_1^{E_6}(\tau,\vecmu)\Eqn{:=}
 \eta(\tau)^8
 \left(\Theta_{E_6}^{[1]}(\tau,\vecmu)
      -\Theta_{E_6}^{[6]}(\tau,\vecmu)\right).
\]
This is a Jacobi form of weight 7. If we set $\vecmu=\veczero$,
it vanishes:
\[
D_1^{E_6}(\tau,\veczero)=0.
\]

The remaining Jacobi form of weight two can be constructed
from $D_1^{E_6}$ by the Hecke transformation of order two.
One obtains
\[
D_2^{E_6}(\tau,\vecmu)
 \Eqn{:=}
  D_1^{E_6}(2\tau,2\vecmu)
  +\frac{1}{128}\sum_{k=0}^1
  D_1^{E_6}\left(\frac{\tau+k}{2},\vecmu\right).
\]

To summarize, we now have seven Jacobi forms
\[\label{E6ABCD}
A_m^{E_6}\ \ (m=1,2,3),\quad
B_2^{E_6},\quad
C_1^{E_6},\quad
D_m^{E_6}\ \ (m=1,2),
\]
which are of weight $4,6,6,7$ respectively and
index given by their subscripts.
We checked that they are independent.
We also checked that $A_m^{E_6},B_2^{E_6},C_1^{E_6}$
are holomorphic Jacobi forms, while $D_m^{E_6}$
are Jacobi cusp forms.
Note that
\[
A_m(\tau,\veczero)=E_4(\tau),\qquad
B_2(\tau,\veczero)=C_1(\tau,\veczero)=E_6(\tau),\qquad
D_m(\tau,\veczero)=0.
\]
As expected, $C_2^{E_6},B_3^{E_6},B_4^{E_6}$ are expressed
as polynomials in these Jacobi forms.
We will use these relations to obtain the results
in the next section.
Again, we do not present concrete expressions here,
as these relations can easily be restored
from the results we will obtain there.

We summarize our choice of independent $E_n$ Jacobi forms
in Table~\ref{gentable}.

\begin{table}[t]
\begin{align}
E_8:\quad&
\begin{array}{|c||c|c|c|c|c|c|}
\hline
\mbox{weight 4}&A_1&A_2&A_3&A_4&A_5&   \\ \hline
\mbox{weight 6}&   &B_2&B_3&B_4&   &B_6\\ \hline
\end{array}\nn\\
E_7:\quad&
\begin{array}{|c||c|c|c|c|}
\hline
\mbox{weight 4}&A_1&A_2&A_3&   \\ \hline
\mbox{weight 6}&   &B_2&B_3&B_4\\
               &C_1&C_2&   &   \\ \hline
\end{array}\nn\\
E_6:\quad&
\begin{array}{|c||c|c|c|c|}
\hline
\mbox{weight 4}&A_1&A_2&A_3\\ \hline
\mbox{weight 6}&C_1&B_2&   \\ \hline
\mbox{weight 7}&\!D_1\!&\!D_2\!&   \\ \hline
\end{array}\nn
\end{align}
\caption{Our choice of independent $E_n$ Jacobi forms.
The subscripts of the Jacobi forms represent their index.
\label{gentable}}
\end{table}
%

\section{Seiberg--Witten curves and generators of weak Jacobi forms}

\subsection{Generalities}

In \cite{Wirthmuller} Wirthm\"uller proved that
for any irreducible root system $R$
excluding $E_8$, the algebra of $W(R)$-invariant Jacobi forms over
the algebra of modular forms $\bbC[E_4,E_6]$
is generated as the polynomial algebra in some
$W(R)$-invariant Jacobi forms
\[\label{Rgenset}
\{\gen_{k(j),m(j)}(\tau,\vecmu)\}\qquad (j=0,1,\ldots,n).
\]
Here, $\{k(j)\}$ and $\{m(j)\}$ are given respectively
by the list of degrees of independent Casimir invariants
of $R$ and the list of
levels of the fundamental representations of
the affine $R$ Lie algebra.
In what follows we explicitly construct $\{\gen_{k(j),m(j)}\}$
for $R=E_6,E_7$
exploiting the Seiberg--Witten curve
for the E-string theory.
We also present a similar set of
meromorphic functions (i.e.~not exactly Jacobi forms) for $R=E_8$.

\subsection{$E_8$ case}

In \cite{Sakai:2011xg}
the Seiberg--Witten curve for the E-string theory \cite{Eguchi:2002fc}
was expressed in terms of the nine
Jacobi forms $A_m,B_m$
given in (\ref{E8AB}). The result is as follows:
\[\label{E8curve}
y^2\Eqn{=}
 4x^3-\frac{1}{12}E_4u^4x-\frac{1}{216}E_6u^6\nn\\
&&-\sum_{m=2}^4\gen_{4-6m,m}u^{4-m}x
  -\sum_{m=1}^6\gen_{6-6m,m}u^{6-m},
\]
where
\[
\gen_{0,1}\Eqn{=}-\frac{4}{E_4}A_1,\nn\\
\gen_{-6,2}\Eqn{=}
  \frac{5}{6E_4^2\Delta}\Bigl(E_4^2B_2-E_6A_1^2\Bigr),\qquad
\gen_{-8,2}=
  \frac{6}{E_4\Delta}\Bigl(-E_4A_2+A_1^2\Bigr),\nn\\
\gen_{-12,3}\Eqn{=}\frac{1}{108E_4^3\Delta^2}
  \Bigl(-7E_4^5A_3-20E_4^3E_6B_3\nn\\
&&\hspace{1.0em}
  -9E_4^4A_1A_2+30E_4^2E_6A_1B_2+(16E_4^3-10E_6^2)A_1^3\Bigr),\nn\\
\gen_{-14,3}\Eqn{=}\frac{1}{9E_4^2\Delta^2}
  \Bigl(-7E_4^2E_6A_3-20E_4^3B_3
       -9E_4E_6A_1A_2+30E_4^2A_1B_2+6E_6A_1^3\Bigr),\nn\\
\gen_{-18,4}\Eqn{=}\frac{1}{1728E_4^4\Delta^3}
  \Bigl((-5E_4^7+5E_4^4E_6^2)B_4
  +(80E_4^6-80E_4^3E_6^2)A_1B_3\nn\\
&&\hspace{1.0em}
  +9E_4^5E_6A_2^2+30E_4^6A_2B_2+25E_4^4E_6B_2^2
  -48E_4^4E_6A_1^2A_2\nn\\
&&\hspace{1.0em}
  +(-140E_4^5+60E_4^2E_6^2)A_1^2B_2
  +(74E_4^3E_6-10E_6^3)A_1^4\Bigr),\nn\\
\gen_{-20,4}\Eqn{=}\frac{1}{864E_4^3\Delta^3}
  \Bigl((E_4^6-E_4^3E_6^2)A_4+(56E_4^5-56E_4^2E_6^2)A_1A_3
  -27E_4^5A_2^2\nn\\
&&\hspace{1.0em}
  -90E_4^3E_6A_2B_2-75E_4^4B_2^2+(180E_4^4-36E_4E_6^2)A_1^2A_2\nn\\
&&\hspace{1.0em}
  +240E_4^2E_6A_1^2B_2
  +(-210E_4^3+18E_6^2)A_1^4\Bigr),\nn\\
\gen_{-24,5}\Eqn{=}\frac{1}{72E_4^5\Delta^3}\Bigl(
  (-21E_4^7+21E_4^4E_6^2)A_5-294E_4^6A_2A_3-770E_4^4E_6B_2A_3\nn\\
&&\hspace{1.0em}
  -840E_4^4E_6A_2B_3-2200E_4^5B_2B_3+168E_4^5A_1^2A_3
  +480E_4^3E_6A_1^2B_3\nn\\
&&\hspace{1.0em}
  -621E_4^5A_1A_2^2+3525E_4^4A_1B_2^2
  +1224E_4^4A_1^3A_2-240E_4^2E_6A_1^3B_2\nn\\
&&\hspace{1.0em}
  +(-456E_4^3+24E_6^2)A_1^5
\Bigr),\nn\\
\noalign{\break}
\gen_{-30,6}\Eqn{=}\frac{1}{13436928E_4^6\Delta^5}\Bigl(
  (-20E_4^{12}+40E_4^9E_6^2-20E_4^6E_6^4)B_6\nn\\
&&\hspace{1.0em}
 +(-189E_4^{10}E_6+378E_4^7E_6^3-189E_4^4E_6^5)A_1A_5\nn\\
&&\hspace{1.0em}
 +(-9E_4^{10}E_6+9E_4^7E_6^3)A_2A_4
 +(-15E_4^{11}+15E_4^8E_6^2)B_2A_4\nn\\
&&\hspace{1.0em}
 +(-180E_4^{11}+180E_4^8E_6^2)A_2B_4
 +(-300E_4^9E_6+300E_4^6E_6^3)B_2B_4\nn\\
&&\hspace{1.0em}
 +(22E_4^9E_6-22E_4^6E_6^3)A_1^2A_4
 +(150E_4^{10}+120E_4^7E_6^2-270E_4^4E_6^4)A_1^2B_4\nn\\
&&\hspace{1.0em}
 +(196E_4^{10}E_6-196E_4^7E_6^3)A_3^2
 +(1120E_4^{11}-1120E_4^8E_6^2)A_3B_3\nn\\
&&\hspace{1.0em}
 +(1600E_4^9E_6-1600E_4^6E_6^3)B_3^2
 +(-2982E_4^9E_6+2982E_4^6E_6^3)A_1A_2A_3\nn\\
&&\hspace{1.0em}
 +(-2520E_4^{10}-4410E_4^7E_6^2+6930E_4^4E_6^4)A_1B_2A_3\nn\\
&&\hspace{1.0em}
 +(3360E_4^{10}-10920E_4^7E_6^2+7560E_4^4E_6^4)A_1A_2B_3\nn\\
&&\hspace{1.0em}
 +(-19800E_4^8E_6+19800E_4^5E_6^3)A_1B_2B_3
 +(2016E_4^8E_6-2016E_4^5E_6^3)A_1^3A_3\nn\\
&&\hspace{1.0em}
 +(-5920E_4^9+7360E_4^6E_6^2-1440E_4^3E_6^4)A_1^3B_3
 +(405E_4^9E_6+162E_4^6E_6^3)A_2^3\nn\\
&&\hspace{1.0em}
 +(1215E_4^{10}+1620E_4^7E_6^2)A_2^2B_2
 +4725E_4^8E_6A_2B_2^2\nn\\
&&\hspace{1.0em}
 +(1125E_4^9+1500E_4^6E_6^2)B_2^3
 +(-9477E_4^8E_6+5103E_4^5E_6^3)A_1^2A_2^2\nn\\
&&\hspace{1.0em}
 +(-9180E_4^9-5400E_4^6E_6^2)A_1^2A_2B_2
 +(20925E_4^7E_6-33075E_4^4E_6^3)A_1^2B_2^2\nn\\
&&\hspace{1.0em}
 +(20304E_4^7E_6-9072E_4^4E_6^3)A_1^4A_2\nn\\
&&\hspace{1.0em}
 +(12780E_4^8+5400E_4^5E_6^2+540E_4^2E_6^4)A_1^4B_2\nn\\
&&\hspace{1.0em}
 +(-11076E_4^6E_6+1512E_4^3E_6^3-36E_6^5)A_1^6
\Bigr).
\]

Since $E_4(e^{2\pi i/3})=0$,
it is very likely that the above $\gen_{k,m}$ have a pole
at $\tau=e^{2\pi i/3}$.
Apart from this flaw, 
$\gen_{k,m}$ satisfy all the conditions required for
$W(E_8)$-invariant weak Jacobi forms (of weight $k$ and index $m$):
By construction they satisfy conditions
(\ref{Weylinv})--(\ref{Modularprop}).
It is also obvious that no fractional powers of $q$
appear in their Fourier expansions. It is known \cite{Eguchi:2002fc}
that they are finite at $q=0$ (see Appendix~A for the concrete
expressions). Thus, the condition (\ref{Fourierform})
is also satisfied.

If we set $\vecmu=\veczero$,
all $\gen_{k,m}$ of negative weight vanish:
\[
\gen_{0,1}(\tau,\veczero)\Eqn{=}-4,\nn\\
\gen_{k,m}(\tau,\veczero)\Eqn{=}0\quad(k< 0).
\]

Although Wirthm\"uller's theorem \cite{Wirthmuller}
does not cover the case of $R=E_8$ and the above
$\gen_{k,m}$ are not exactly Jacobi forms,
it would still be interesting to examine
to what extent
the statements of the theorem hold
for $R=E_8$.\footnote{
There is an algebro-geometric explanation
why $E_8$ should be exceptional \cite{Friedman:1997yq}.}
Interestingly, there is a small mismatch between
the above $\gen_{k,m}$ and the generators that would be expected
supposing Wirthm\"uller's theorem held:
The theorem would require a generator of weight $-2$ and index $2$
instead of $\gen_{-6,2}$.
In fact such a Jacobi form can easily be constructed as
\[
\tilde\gen_{-2,2}:=E_4\gen_{-6,2}.
\]
However, if one replaces $\gen_{-6,2}$ with $\tilde\gen_{-2,2}$
in the generator set, certain $W(E_8)$-invariant
Jacobi forms cannot be generated
over the ring of modular forms $\bbC[E_4,E_6]$.\footnote{
The author is grateful to Haowu Wang for explaining
this point and also indicating some misunderstandings
about $W(E_8)$-invariant Jacobi forms in the previous manuscript.}

Though the above $\gen_{k,m}$ themselves are not exactly Jacobi forms,
one can still consider the polynomial algebra
generated by $\gen_{k,m}$ over $\bbC[E_4,E_6]$.
To the best of our knowledge, this algebra 
seems general enough to contain all $E_8$
weak Jacobi forms whose concrete expressions are known.
Therefore we conjecture that
the algebra of $W(E_8)$-invariant weak Jacobi forms
would be a proper subset of the polynomial algebra generated by 
$\gen_{k,m}$ over $\bbC[E_4,E_6]$.
It would be very interesting to investigate this problem 
in a more mathematically rigorous manner.

\subsection{$E_7$ case}

One can reduce the Seiberg--Witten curve
presented in the last subsection to the curve
that has only $W(E_7)$ symmetry
by setting $\vecmu=\vecmu^{(7)}$.
The curve can be expressed in terms of
the eight $E_7$ Jacobi forms constructed in section~2.3.
It is expected that the elliptic fibration described by
this curve develops a degenerate fiber.
(It was systematically studied in \cite{Eguchi:2002nx}
how special values of $\vecmu$
correspond to degenerations of the elliptic fibration
described by the Seiberg--Witten curve for the E-string theory.)
One immediate outcome of expressing the curve in
terms of the eight $E_7$ Jacobi forms is that
one can directly see this fiber degeneration as the factorization
of the discriminant.
For the elliptic curve in the Weierstrass form
\[\label{Weierstrass}
y^2=4x^3-fx-g,
\]
the discriminant is given by
\[\label{Weierdisc}
D = f^3-27g^2.
\]
For the above Seiberg--Witten curve
expressed in terms of the eight $E_7$ Jacobi forms,
the discriminant indeed factorizes as
\[\label{discE7a}
D=(u-u_0)^2P_{10}(u),
\]
where
\[\label{u0}
u_0 = \frac{E_4C_1-E_6A_1}{12E_4\Delta}
\]
and $P_{10}(u)$ is some tenth degree polynomial in $u$.
(In this subsection we omit superscript $E_7$ from Jacobi forms.)
This is not the only peculiar feature
of the above reduced curve.
We can in fact transform the curve into the form
of the general deformation of a singularity of type
$\tilde{E}_7$, as we will see below.

The general deformation of a singularity of type
$\tilde{E}_7$ takes the form \cite{Friedman:1997yq}
\[\label{newE7curve}
y^2\Eqn{=}
 4ux^3-\frac{1}{12}E_4u^3x-\frac{1}{216}E_6u^4\nn\\
&&+\gen_{0,1}u^3+\gen_{-2,1}u^2x
  +\gen_{-6,2}u^2+\gen_{-8,2}ux+\gen_{-10,2}x^2\nn\\
&&+\gen_{-12,3}u+\gen_{-14,3}x
  +\gen_{-18,4}.
\]
For the moment $\gen_{k,m}$ are just deformation parameters.
We formally assign weights $-6,-4,-9,k$ and indices $1,1,2,m$
to $u,x,y,\gen_{k,m}$ respectively,
so that all terms in the equation are
of weight $-18$ and index 4.
The elliptic curve (\ref{newE7curve})
can be transformed into the Weierstrass form (\ref{Weierstrass})
with
\[\label{Weiercoeffs}
f=\frac{1}{12}E_4u^4+\sum_{k=1}^4f_ku^{4-k},\qquad
g=\frac{1}{216}E_6u^6+\sum_{k=1}^6g_ku^{6-k}
\]
in the following manner: First, perform a translation of $x$
to remove the quadratic term in $x$.
Next, rescale the variables as
$x\to u^{-1}x,\ y\to u^{-1}y$.
We then obtain the Weierstrass from with $f,g$ being
of the form (\ref{Weiercoeffs}).
One finds that the discriminant of this curve
factorizes as
\[\label{discE7b}
D=u^2\tilde{P}_{10}(u),
\]
where $\tilde{P}_{10}(u)$ is some tenth degree polynomial in $u$.
Another peculiar feature of the curve is that
the coefficients $f_4,\,g_6$ also factorize
\[\label{ta4tb6}
f_4=\frac{(\gen_{-10,2})^2}{12},\qquad
g_6=-\frac{(\gen_{-10,2})^3}{216}.
\]
The locations of the double roots of the discriminants
(\ref{discE7a}) and (\ref{discE7b})
imply that
the original Seiberg--Witten curve with $\vecmu=\vecmu^{(7)}$
is identified with
the above obtained curve in the Weierstrass form
by the translation $u\to u+u_0$.
Indeed, after the translation is applied to the former curve,
one can see the factorizations of coefficients
as in (\ref{ta4tb6}) and determine $\gen_{-10,2}$.
Furthermore, by comparing two curves term by term,
one can fully determine the coefficients $\gen_{k,m}$
in (\ref{newE7curve}).
The results are as follows:
\[
\gen_{0,1}\Eqn{=}
 \frac{E_4^2A_1-E_6C_1}{432\Delta},\qquad
\gen_{-2,1}=
 \frac{E_6A_1-E_4C_1}{36\Delta},\nn\\
\gen_{-6,2}\Eqn{=}
 \frac{1}{82944\Delta^2}\left(
 (-25E_4^3+25E_6^2)B_2+(-11E_4^3+11E_6^2)C_2\right.\nn\\
&&\hspace{1.5em}\left.
 -36E_4E_6A_1^2+72E_4^2A_1C_1-36E_6C_1^2\right),\nn\\
\gen_{-8,2}\Eqn{=}
 \frac{1}{288\Delta^2}\left(
 (E_4^3-E_6^2)A_2-E_4^2A_1^2+2E_6A_1C_1-E_4C_1^2\right),\nn\\
\gen_{-10,2}\Eqn{=}
 \frac{1}{576E_4\Delta^2}\left(
 (-15E_4^3+15E_6^2)B_2+(11E_4^3-11E_6^2)C_2\right.\nn\\
&&\hspace{1.5em}\left.
 -4E_4E_6A_1^2+8E_4^2A_1C_1-4E_6C_1^2\right),\nn\\
\gen_{-12,3}\Eqn{=}
 \frac{1}{746496E_4\Delta^3}\left(
 (28E_4^6-28E_4^3E_6^2)A_3+(80E_4^4E_6-80E_4E_6^3)B_3\right.\nn\\
&&\hspace{1.5em}
 +(36E_4^5-36E_4^2E_6^2)A_1A_2+(-45E_4^3E_6+45E_6^3)A_1B_2\nn\\
&&\hspace{1.5em}
 +(-75E_4^4+75E_4E_6^2)C_1B_2+(33E_4^3E_6-33E_6^3)A_1C_2\nn\\
&&\hspace{1.5em}
 +(-33E_4^4+33E_4E_6^2)C_1C_2+(-64E_4^4+92E_4E_6^2)A_1^3\nn\\
&&\hspace{1.5em}\left.
 -84E_4^2E_6A_1^2C_1+(96E_4^3-12E_6^2)A_1C_1^2
 -28E_4E_6C_1^3\right),\nn\\
\gen_{-14,3}\Eqn{=}
 \frac{1}{15552\Delta^3}\left(
 (7E_4^3E_6-7E_6^3)A_3+(20E_4^4-20E_4E_6^2)B_3\right.\nn\\
&&\hspace{1.5em}
 +(9E_4^3-9E_6^2)C_1A_2
 +(-30E_4^3+30E_6^2)A_1B_2\nn\\
&&\hspace{1.5em}\left.
 +3E_4E_6A_1^3-9E_4^2A_1^2C_1+9E_6A_1C_1^2-3E_4C_1^3\right),\nn\\
\gen_{-18,4}\Eqn{=}
 \frac{1}{5971968E_4\Delta^4}\left(
 (10E_4^7-20E_4^4E_6^2+10E_4E_6^4)B_4\right.\nn\\
&&\hspace{1.5em}
 +(-56E_4^5E_6+56E_4^2E_6^3)A_1A_3
 +(56E_4^6-56E_4^3E_6^2)C_1A_3\nn\\
&&\hspace{1.5em}
 +(-160E_4^6+160E_4^3E_6^2)A_1B_3
 +(160E_4^4E_6-160E_4E_6^3)C_1B_3\nn\\
&&\hspace{1.5em}
 +(-18E_4^5E_6+18E_4^2E_6^3)A_2^2
 +(-105E_4^6+150E_4^3E_6^2-45E_6^4)A_2B_2\nn\\
&&\hspace{1.5em}
 +(33E_4^6-66E_4^3E_6^2+33E_6^4)A_2C_2
 +(-50E_4^4E_6+50E_4E_6^3)B_2^2\nn\\
&&\hspace{1.5em}
 +(12E_4^4E_6-12E_4E_6^3)A_1^2A_2
 +(96E_4^5-96E_4^2E_6^2)A_1C_1A_2\nn\\
&&\hspace{1.5em}
 +(-12E_4^3E_6+12E_6^3)C_1^2A_2
 +(325E_4^5-325E_4^2E_6^2)A_1^2B_2\nn\\
&&\hspace{1.5em}
 +(-90E_4^3E_6+90E_6^3)A_1C_1B_2
 +(-75E_4^4+75E_4E_6^2)C_1^2B_2\nn\\
&&\hspace{1.5em}
 +(-33E_4^5+33E_4^2E_6^2)A_1^2C_2
 +(66E_4^3E_6-66E_6^3)A_1C_1C_2\nn\\
&&\hspace{1.5em}
 +(-33E_4^4+33E_4E_6^2)C_1^2C_2
 -8E_4^3E_6A_1^4
 +(-152E_4^4+184E_4E_6^2)A_1^3C_1\nn\\
&&\hspace{1.5em}\left.
 -48E_4^2E_6A_1^2C_1^2
 +(56E_4^3-24E_6^2)A_1C_1^3
 -8E_4E_6C_1^4
\right).
\]
Here, $A_m,B_m,C_m$ are the holomorphic Jacobi forms
constructed in section~2.3 and we have omitted superscript $E_7$.

In contrast to the $E_8$ case, 
the above $\gen_{k,m}$ are genuine $W(E_7)$-invariant weak Jacobi forms
(of weight $k$ and index $m$).
This can be shown as follows:
By construction they satisfy conditions
(\ref{Weylinv})--(\ref{Modularprop})
and no fractional powers of $q$
appear in their Fourier expansions.
We checked explicitly that they are finite at $q=0$.
We present the concrete expressions of $\gen_{k,m}$
at $q=0$ in Appendix~A.
On the other hand,
it is less trivial to show
that $\gen_{k,m}$ are holomorphic in $\tau$.
As the expressions of
$\gen_{-10,2}, \gen_{-12,3}$ and $\gen_{-18,4}$ 
contain $E_4$ in the denominator, these generators
may have a pole at $\tau=e^{2\pi i/3}$.
By carefully examining the structure of these expressions,
one finds that these generators can be written as
\[
\gen_{-10,2}\Eqn{=}
 \frac{E_6X+\cdots}{576\Delta^2},\nn\\
\gen_{-12,3}\Eqn{=}
 \frac{3E_6^2A_1X+\cdots}{746496\Delta^3},\nn\\
\gen_{-18,4}\Eqn{=}
 \frac{E_6^2(-E_6A_2+2A_1C_1)X+\cdots}{1990656\Delta^4},
\]
where ``$\cdots$'' are some polynomials in
$A_m,B_m,C_m,E_k$ and
\[
X:=\frac{15E_6B_2-11E_6C_2-4C_1^2}{E_4}.
\]
Clearly, potential divergence can arise only through $X$.
Therefore the proof boils down to showing that $X$
is regular at $\tau=e^{2\pi i/3}$.
This can be done as follows:
The relation (\ref{E7A4rel}) can be rewritten as
\[
3X^2-24A_1^2X
\Eqn{=}-13824\Delta A_4
  -448E_4^2A_1A_3+448E_6C_1A_3-1280E_6A_1B_3\nn\\
&&+1280E_4C_1B_3+216E_4^2A_2^2-1440E_4A_1^2A_2+720E_6A_2B_2\nn\\
&&+288C_1^2A_2+1275E_4B_2^2-990E_4B_2C_2-2640A_1C_1B_2\nn\\
&&+363E_4C_2^2+528A_1C_1C_2+1680A_1^4.
\]
Since the right-hand side is holomorphic in $\tau$,
$X$ has to be regular at $\tau=e^{2\pi i/3}$.
Hence, we have shown that all
$\gen_{k,m}$ are indeed $W(E_7)$-invariant weak Jacobi forms.

The above $\gen_{k,m}$ satisfy
all the conditions required
for the generators in the Wirthm\"uller's theorem
explained in section~3.1.
Thus we conclude that they give a full set of
generators of the algebra of $W(E_7)$-invariant weak Jacobi forms
over the algebra of modular forms $\bbC[E_4,E_6]$.

If we set $\vecmu=\veczero$, the generators become
\[\label{mlgenE7}
\gen_{0,1}(\tau,\veczero)\Eqn{=}4,\nn\\
\gen_{k,m}(\tau,\veczero)\Eqn{=}0\quad(k< 0).
\]
%

\subsection{$E_6$ case}

In the same way as in the $E_7$ case, one can reduce
the Seiberg--Witten curve for the E-string theory
to the curve that has only $W(E_6)$ symmetry
and transform it into the form of the deformed
singularity of type $\tilde{E}_6$.

The general deformation of a singularity of type
$\tilde{E}_6$ takes the form \cite{Friedman:1997yq}
\[\label{newE6curve}
uy^2\Eqn{=}
 4x^3-\frac{1}{12}E_4u^2x-\frac{1}{216}E_6u^3\nn\\
&&+\gen_{0,1}u^2+\gen_{-2,1}ux+\gen_{-5,1}xy
  +\gen_{-6,2}u+\gen_{-8,2}x+\gen_{-9,2}y
  +\gen_{-12,3}.\quad
\]
One can formally assign weights $-6,-4,-3,k$ and indices $1,1,1,m$
to $u,x,y,\gen_{k,m}$ respectively,
so that all terms in the equation are of weight $-12$ and index 3.
The curve (\ref{newE6curve}) can be transformed
into the Weierstrass form
(\ref{Weierstrass}) with (\ref{Weiercoeffs})
as follows:
First, perform a translation of $y$ to
eliminate the linear terms in $y$.
Next, perform a translation of $x$ to
eliminate the quadratic terms in $x$.
Finally, rescale the variables as
$x\to u^{-1}x,\ y\to u^{-2}y$.

Next, we reduce the original Seiberg--Witten curve in section~3.2:
We first set $\vecmu=\vecmu^{(6)}$, then rewrite it in terms
of the seven $E_6$ Jacobi forms constructed in section~2.4,
and finally replace $u$ by $u+u_0$.
Here, $u_0$ is given in (\ref{u0}).
By comparing this curve with the above curve in the Weierstrass form,
we are able to determine
all the coefficients $\gen_{k,m}$ in (\ref{newE6curve}).
The results are as follows:
\[\label{genE6}
\gen_{0,1}\Eqn{=}
 \frac{E_4^2A_1-E_6C_1}{432\Delta},\qquad
\gen_{-2,1}=
 \frac{E_6A_1-E_4C_1}{36\Delta},\qquad
\gen_{-5,1}=
 \frac{2iD_1}{\Delta},\nn\\
\gen_{-6,2}\Eqn{=}
 \frac{1}{10368\Delta^2}\left(
 (-5E_4^3+5E_6^2)B_2\right.\nn\\
&&\hspace{1.5em}\left.
 -5E_4E_6A_1^2+10E_4^2A_1C_1-5E_6C_1^2+72E_4D_1^2\right),\nn\\
\gen_{-8,2}\Eqn{=}
 \frac{1}{288\Delta^2}\left(
 (E_4^3-E_6^2)A_2-E_4^2A_1^2+2E_6A_1C_1-E_4C_1^2\right),\nn\\
\gen_{-9,2}\Eqn{=}
 \frac{i}{108E_4\Delta^2}\left(
 (-8E_4^3+8E_6^2)D_2-3E_4^2A_1D_1+3E_6C_1D_1\right),\nn\\
\gen_{-12,3}\Eqn{=}
 \frac{1}{186624E_4^2\Delta^3}\left(
 (7E_4^7-14E_4^4E_6^2+7E_4E_6^4)A_3
 +(9E_4^6-9E_4^3E_6^2)A_1A_2\right.\nn\\
&&\hspace{1.5em}
 +(-9E_4^4E_6+9E_4E_6^3)C_1A_2
 +(30E_4^4E_6-30E_4E_6^3)A_1B_2\nn\\
&&\hspace{1.5em}
 +(-30E_4^5+30E_4^2E_6^2)C_1B_2
 +(1152E_4^3E_6-1152E_6^3)D_1D_2\nn\\
&&\hspace{1.5em}
 +(-16E_4^5+23E_4^2E_6^2)A_1^3
 -21E_6E_4^3A_1^2C_1
 +(30E_4^4-9E_4E_6^2)A_1C_1^2\nn\\
&&\hspace{1.5em}\left.
 -7E_4^2E_6C_1^3
 +(432E_4^3-432E_6^2)C_1D_1^2
\right).
\]
Here, $A_m,B_2,C_1,D_m$ are the Jacobi forms
constructed in section~2.4 and we have omitted superscript $E_6$.

When $\vecmu=\vecmu^{(6)}$,
$\gen_{k,m}^{E_7}$ are expressed as
polynomials in $\gen_{k,m}^{E_6}$.
The relations are extremely simple:
\begin{align}\label{genrelE7E6}
\gen_{0,1}^{E_7}&=\gen_{0,1}^{E_6},&
\gen_{-2,1}^{E_7}&=\gen_{-2,1}^{E_6},&
\gen_{-6,2}^{E_7}&=\gen_{-6,2}^{E_6},\nn\\
\gen_{-8,2}^{E_7}&=\gen_{-8,2}^{E_6},&
\gen_{-10,2}^{E_7}&=
 \frac{1}{4}\left(\gen_{-5,1}^{E_6}\right)^2,&
\gen_{-12,3}^{E_7}&=\gen_{-12,3}^{E_6},\nn\\
\gen_{-14,3}^{E_7}&=
 \frac{1}{2}\gen_{-5,1}^{E_6}\gen_{-9,2}^{E_6},&
\gen_{-18,4}^{E_7}&=
 \frac{1}{4}\left(\gen_{-9,2}^{E_6}\right)^2.
\end{align}
This is in agreement with the description of
the generators of $E_7$ Jacobi forms in \cite{Wirthmuller}.

In the same way as in the $E_7$ case,
one can show that
$\gen_{k,m}$ given in (\ref{genE6})
are genuine $W(E_6)$-invariant weak Jacobi forms
(of weight $k$ and index $m$).
We present the concrete expressions of them at $q=0$ in Appendix~A.
The expressions of $\gen_{-9,2}^{E_6}$ and $\gen_{-12,3}^{E_6}$
contain $E_4$ in the denominator and thus they may
have a pole at $\tau=e^{2\pi i/3}$. However,
since all $\gen_{k,m}^{E_7}$ are genuine Jacobi forms,
it is clear from (\ref{genrelE7E6}) that
$\gen_{-9,2}^{E_6}$ and $\gen_{-12,3}^{E_6}$ are
in fact regular at $\tau=e^{2\pi i/3}$.

The above $\gen_{k,m}$ satisfy
all the conditions required
for the generators in the Wirthm\"uller's theorem
explained in section~3.1.
Thus we conclude that they give a full set of
generators of the algebra of $W(E_6)$-invariant weak Jacobi forms
over the algebra of modular forms $\bbC[E_4,E_6]$.

If we set $\vecmu=\veczero$, the generators become
\[\label{mlgenE6}
\gen_{0,1}(\tau,\veczero)\Eqn{=}4,\nn\\
\gen_{k,m}(\tau,\veczero)\Eqn{=}0\quad(k< 0).
\]

In \cite{Satake:1993cp} $E_6$ Jacobi forms were used
in the study of the flat structure for the elliptic singularity
of type $\tilde{E}_6$.
The generators specified in \cite{Satake:1993cp}
(up to the overall factor $e(-mt)$)
are expressed in terms of our generators as
\begin{align}
\varphi_0&=18\gen_{0,1},&
\varphi_1&=\frac{3}{2}\gen_{-2,1},&
\varphi_2&=-\frac{i}{2}\gen_{-5,1},\nn\\
\varphi_3&=-9\gen_{-6,2}-\frac{5}{64}E_4\left(\gen_{-5,1}\right)^2,&
\varphi_4&=-3\gen_{-8,2},&
\varphi_5&=-3i\gen_{-9,2},\nn\\
\varphi_6&=27\gen_{-12,3}
  -\frac{159}{16}\gen_{-2,1}\left(\gen_{-5,1}\right)^2.\hspace{-3em}
\end{align}
%

\vspace{3ex}

\begin{center}
  {\bf Acknowledgments}
\end{center}

The author would like to thank Jie Gu for stimulating questions,
which inspired him to start this work.
The author would also like to thank Haowu Wang for
valuable comments and discussions.
This work was supported in part by
JSPS KAKENHI Grant Number 26400257
and JSPS Japan--Russia Research Cooperative Program.



\appendix

\section{Seiberg--Witten curves at $q=0$}

In this appendix we present the Seiberg--Witten curves of type
$\tilde{E}_n$ at $q=0$ ($\tau=i\infty$).
These Seiberg--Witten curves describe the low-energy theory of 5d
$\grp{SU}(2)\ N_{\rm f}=7$ gauge theory on $\bbR^4\times S^1$.
The $\tilde{E}_8$ curve below is
the 5d $E_8$ curve in \cite{Eguchi:2002fc}.
The $\tilde{E}_n\ (n=7,6)$ curves below are not equivalent to
the 5d $E_n$ curves in \cite{Eguchi:2002fc}:
The former curves give a degenerate
fiber at $u=0$ while the latter curves
give a degenerate fiber at $u=\infty$.
Physically, the former curves describe special cases of
5d $\grp{SU}(2)\ N_{\rm f}=7$ theory while the latter ones
describe 5d $\grp{SU}(2)\ N_{\rm f}=n-1$ theories.

For each $E_n$ let us define
\[
\gen_{k,m}^{(0)}(\vecmu)
 :=\gen_{k,m}(\tau=i\infty,\vecmu)
\]
and the
Weyl orbit character associated with the fundamental weight $\vecL_j$
\[
w_j(\vecmu):= \sum_{\vecv\in{\cal O}\left(\vecL_j\right)}
e^{2\pi i\vecv\cdot\vecmu}.
\]

\noindent
$\bullet$
$\tilde{E}_8$ curve:
\[
y^2\Eqn{=}
 4x^3-\frac{1}{12}u^4x-\frac{1}{216}u^6\nn\\
&&-\sum_{m=2}^4\gen_{4-6m,m}^{(0)}u^{4-m}x
  -\sum_{m=1}^6\gen_{6-6m,m}^{(0)}u^{6-m},
\]
where
\[
\gen_{0,1}^{(0)}
\Eqn{=}-4,\qquad
\gen_{-6,2}^{(0)}
 =-\frac{1}{18}w_1-3w_8+840,\qquad
\gen_{-8,2}^{(0)}
 =-\frac{2}{3}w_1+12w_8-1440,\nn\\
\gen_{-12,3}^{(0)}
\Eqn{=}-\frac{1}{6}w_2-4w_7-8w_1+528w_8-79680,\nn\\
\gen_{-14,3}^{(0)}
\Eqn{=}-2w_2+96w_1-1152w_8+103680,\nn\\
\gen_{-18,4}^{(0)}
\Eqn{=}\frac{2}{9}w_1^2-\frac{1}{3}w_3-\frac{16}{3}w_6
  -24w_1w_8-120w_8^2\nn\\
&&+\frac{424}{3}w_2+1272w_7+4608w_1-25920w_8+3939840,\nn\\
\gen_{-20,4}^{(0)}
\Eqn{=}\frac{4}{3}w_1^2-4w_3-16w_6-48w_1w_8-144w_8^2\nn\\
&&+400w_2+1440w_7+1728w_1+41472w_8-2073600,\nn\\
\gen_{-24,5}^{(0)}
\Eqn{=}\frac{2}{3}w_1w_2-4w_5-16w_1w_7+64w_2w_8+288w_7w_8
  -96w_1^2-60w_3-160w_6\nn\\
&&+3456w_8^2+800w_2-24480w_7-108480w_1+933120w_8-97873920,\nn\\
\noalign{\break}
\gen_{-30,6}^{(0)}
\Eqn{=}-\frac{8}{27}w_1^3+w_2^2+\frac{4}{3}w_1w_3
  -4w_4-\frac{32}{3}w_1w_6-48w_1^2w_8+48w_2w_7+288w_7^2\nn\\
&&-40w_3w_8-480w_6w_8-2592w_1w_8^2-9792w_8^3
  +\frac{1124}{3}w_1w_2+548w_5\nn\\
&&+6688w_1w_7+1884w_2w_8+25632w_7w_8+24576w_1^2
  +12920w_3+88320w_6\nn\\
&&+578688w_1w_8+1714176w_8^2-1694400w_2-8460000w_7-30102720w_1\nn\\
&&-104198400w_8+721612800.
\]

\noindent
$\bullet$
$\tilde{E}_7$ curve:
\[
y^2\Eqn{=}
 4ux^3-\frac{1}{12}u^3x-\frac{1}{216}u^4\nn\\
&&+\gen_{0,1}^{(0)}u^3+\gen_{-2,1}^{(0)}u^2x
  +\gen_{-6,2}^{(0)}u^2+\gen_{-8,2}^{(0)}ux+\gen_{-10,2}^{(0)}x^2\nn\\
&&+\gen_{-12,3}^{(0)}u+\gen_{-14,3}^{(0)}x
  +\gen_{-18,4}^{(0)},
\]
where
\[
\gen_{0,1}^{(0)}\Eqn{=}
 \frac{1}{36}w_7+\frac{22}{9},\qquad
\gen_{-2,1}^{(0)}=
 \frac{1}{3}w_7-\frac{56}{3},\nn\\
\gen_{-6,2}^{(0)}\Eqn{=}
 -\frac{1}{16}w_7^2+\frac{26}{9}w_1+\frac{5}{36}w_2+\frac{1}{36}w_6
 -6w_7+67,\nn\\
\gen_{-8,2}^{(0)}\Eqn{=}
 -\frac{1}{2}w_7^2-\frac{32}{3}w_1+\frac{4}{3}w_2
 +\frac{2}{3}w_6+32w_7-152,\nn\\
\gen_{-10,2}^{(0)}\Eqn{=}
 -w_7^2+32w_1-4w_2+4w_6-32w_7+176,\nn\\
\gen_{-12,3}^{(0)}\Eqn{=}
 \frac{7}{108}w_7^3-\frac{32}{9}w_1w_7-\frac{7}{18}w_2w_7
 -\frac{1}{9}w_6w_7+\frac{46}{9}w_7^2\nn\\
&&
-\frac{1244}{9}w_7
 +\frac{736}{9}w_1-\frac{77}{9}w_2+\frac{10}{3}w_3+\frac{1}{3}w_5
 +\frac{104}{9}w_6+\frac{13216}{27},\nn\\
\gen_{-14,3}^{(0)}\Eqn{=}
  \frac{1}{3}w_7^3+\frac{64}{3}w_1w_7-\frac{2}{3}w_2w_7
 -\frac{4}{3}w_6w_7-8w_7^2\nn\\
&&
 -\frac{896}{3}w_1-\frac{116}{3}w_2-8w_3+4w_5
 -\frac{160}{3}w_6-80w_7-\frac{3968}{3},\nn\\
\gen_{-18,4}^{(0)}\Eqn{=}
 -\frac{1}{36}w_7^4-\frac{4}{9}w_1w_7^2+\frac{2}{9}w_2w_7^2
 +\frac{1}{9}w_6w_7^2-\frac{16}{9}w_7^3
 +64w_1^2+16w_1w_6-w_2^2\nn\\
&&
 -\frac{896}{9}w_1w_7-\frac{326}{9}w_2w_7
 -\frac{8}{3}w_3w_7-\frac{2}{3}w_5w_7-\frac{64}{9}w_6w_7
 -\frac{596}{3}w_7^2+\frac{29888}{9}w_1\nn\\
&&
 +\frac{1184}{9}w_2+\frac{208}{3}w_3+4w_4
 -\frac{32}{3}w_5+\frac{5632}{9}w_6+\frac{2816}{9}w_7
 +\frac{111488}{9}.
\]

\noindent
$\bullet$
$\tilde{E}_6$ curve:
\[
uy^2\Eqn{=}
 4x^3-\frac{1}{12}u^2x-\frac{1}{216}u^3\nn\\
&&+\gen_{0,1}^{(0)}u^2+\gen_{-2,1}^{(0)}ux+\gen_{-5,1}^{(0)}xy
  +\gen_{-6,2}^{(0)}u+\gen_{-8,2}^{(0)}x+\gen_{-9,2}^{(0)}y
  +\gen_{-12,3}^{(0)},\quad
\]
where
\[
\gen_{0,1}^{(0)}\Eqn{=}
 \frac{1}{36}w_1+\frac{1}{36}w_6+\frac{5}{2},\quad\ \,
\gen_{-2,1}^{(0)}=
 \frac{1}{3}w_1+\frac{1}{3}w_6-18,\quad\ \,
\gen_{-5,1}^{(0)}=
 i\left(2w_1-2w_6\right),\nn\\
\gen_{-6,2}^{(0)}\Eqn{=}
 -\frac{1}{16}w_1^2-\frac{1}{16}w_6^2
 -\frac{7}{72}w_1w_6+3w_2
 +\frac{1}{6}w_3+\frac{1}{6}w_5
 -\frac{10}{3}w_1-\frac{10}{3}w_6+54,\nn\\
\gen_{-8,2}^{(0)}\Eqn{=}
 -\frac{1}{2}w_1^2-\frac{1}{2}w_6^2
 -\frac{1}{3}w_1w_6-12w_2+2w_3+2w_5+20w_1+20w_6-108,\nn\\
\gen_{-9,2}^{(0)}\Eqn{=}
 i\left(-\frac{1}{3}w_1^2+\frac{1}{3}w_6^2
  +2w_3-2w_5-14w_1+14w_6\right),\nn\\
\gen_{-12,3}^{(0)}\Eqn{=}
 \frac{7}{108}w_1^3+\frac{7}{108}w_6^3
 +\frac{1}{12}w_1^2w_6+\frac{1}{12}w_1w_6^2
 -2w_1w_2-2w_6w_2\nn\\
&&
 -\frac{1}{2}w_1w_3-\frac{1}{2}w_6w_5
 -\frac{1}{6}w_1w_5-\frac{1}{6}w_3w_6
 +\frac{11}{6}w_1^2+\frac{11}{6}w_6^2\nn\\
&&
 +15w_1w_6+4w_4-12w_2-6w_3-6w_5-60w_1-60w_6-72.
\]
%

\section{Simple roots and fundamental weights of $E_n$}

Let $\{\vece_j\}\ (j=1,2,\ldots,8)$ be the orthonormal basis
of $\bbC^8$.

\noindent
$\bullet$
The simple roots of $E_8$:
\[
\vecal_1^{E_8}\Eqn{=}\tfrac{1}{2}\left(
  \vece_1-\vece_2-\vece_3-\vece_4
 -\vece_5-\vece_6-\vece_7+\vece_8
  \right),\nn\\
\vecal_2^{E_8}\Eqn{=}\vece_1+\vece_2,\nn\\
\vecal_j^{E_8}\Eqn{=}-\vece_{j-2}+\vece_{j-1}\quad(j=3,4,\ldots,8).
\]
\noindent
$\bullet$
The fundamental weights of $E_8$:
\[
\vecL_1^{E_8}\Eqn{=}2\vece_8,\nn\\
\vecL_2^{E_8}\Eqn{=}\tfrac{1}{2}\vece_1+\tfrac{1}{2}\vece_2
  +\tfrac{1}{2}\vece_3+\tfrac{1}{2}\vece_4
  +\tfrac{1}{2}\vece_5+\tfrac{1}{2}\vece_6
  +\tfrac{1}{2}\vece_7+\tfrac{5}{2}\vece_8,\nn\\
\vecL_3^{E_8}\Eqn{=}-\tfrac{1}{2}\vece_1+\tfrac{1}{2}\vece_2
  +\tfrac{1}{2}
\vece_3+\tfrac{1}{2}\vece_4+\tfrac{1}{2}\vece_5
  +\tfrac{1}{2}\vece_6+\tfrac{1}{2}\vece_7
  +\tfrac{7}{2}\vece_8,\nn\\
\vecL_4^{E_8}\Eqn{=}\vece_3+\vece_4+\vece_5
  +\vece_6+\vece_7+5\vece_8,\nn\\
\vecL_5^{E_8}\Eqn{=}\vece_4+\vece_5+\vece_6
  +\vece_7+4\vece_8,\nn\\
\vecL_6^{E_8}\Eqn{=}\vece_5+\vece_6+\vece_7+3\vece_8,\nn\\
\vecL_7^{E_8}\Eqn{=}\vece_6+\vece_7+2\vece_8,\nn\\
\vecL_8^{E_8}\Eqn{=}\vece_7+\vece_8.
\]
\newpage
\noindent
$\bullet$
The simple roots of $E_7$:
\[
\vecal_j^{E_7}:=\vecal_j^{E_8}\quad(j=1,2,\ldots,7).
\]
\noindent
$\bullet$
The fundamental weights of $E_7$:
\[
\vecL_1^{E_7}\Eqn{=}-\vece_7+\vece_8,\nn\\
\vecL_2^{E_7}\Eqn{=}
  \tfrac{1}{2}\vece_1+\tfrac{1}{2}\vece_2+\tfrac{1}{2}\vece_3
 +\tfrac{1}{2}\vece_4+\tfrac{1}{2}\vece_5+\tfrac{1}{2}\vece_6
 -\vece_7+\vece_8,\nn\\
\vecL_3^{E_7}\Eqn{=}
 -\tfrac{1}{2}\vece_1+\tfrac{1}{2}\vece_2+\tfrac{1}{2}\vece_3
 +\tfrac{1}{2}\vece_4+\tfrac{1}{2}\vece_5+\tfrac{1}{2}\vece_6
 -\tfrac{3}{2}\vece_7+\tfrac{3}{2}\vece_8,\nn\\
\vecL_4^{E_7}\Eqn{=}
 \vece_3+\vece_4+\vece_5+\vece_6-2\vece_7+2\vece_8,\nn\\
\vecL_5^{E_7}\Eqn{=}
 \vece_4+\vece_5+\vece_6-\tfrac{3}{2}\vece_7+\tfrac{3}{2}\vece_8,\nn\\
\vecL_6^{E_7}\Eqn{=}
 \vece_5+\vece_6-\vece_7+\vece_8,\nn\\
\vecL_7^{E_7}\Eqn{=}
 \vece_6-\tfrac{1}{2}\vece_7+\tfrac{1}{2}\vece_8.
\]
\noindent
$\bullet$
The simple roots of $E_6$:
\[
\vecal_j^{E_6}:=\vecal_j^{E_8}\quad(j=1,2,\ldots,6).
\]
\noindent
$\bullet$
The fundamental weights of $E_6$:
\[
\vecL_1^{E_6}\Eqn{=}
 -\tfrac{2}{3}\vece_6-\tfrac{2}{3}\vece_7+\tfrac{2}{3}\vece_8,\nn\\
\vecL_2^{E_6}\Eqn{=}
  \tfrac{1}{2}\vece_1+\tfrac{1}{2}\vece_2+\tfrac{1}{2}\vece_3
 +\tfrac{1}{2}\vece_4+\tfrac{1}{2}\vece_5-\tfrac{1}{2}\vece_6
 -\tfrac{1}{2}\vece_7+\tfrac{1}{2}\vece_8,\nn\\
\vecL_3^{E_6}\Eqn{=}
 -\tfrac{1}{2}\vece_1+\tfrac{1}{2}\vece_2+\tfrac{1}{2}\vece_3
 +\tfrac{1}{2}\vece_4+\tfrac{1}{2}\vece_5-\tfrac{5}{6}\vece_6
 -\tfrac{5}{6}\vece_7+\tfrac{5}{6}\vece_8,\nn\\
\vecL_4^{E_6}\Eqn{=}
  \vece_3+\vece_4+\vece_5-\vece_6-\vece_7+\vece_8,\nn\\
\vecL_5^{E_6}\Eqn{=}
  \vece_4+\vece_5
 -\tfrac{2}{3}\vece_6-\tfrac{2}{3}\vece_7+\tfrac{2}{3}\vece_8,\nn\\
\vecL_6^{E_6}\Eqn{=}
 \vece_5-\tfrac{1}{3}\vece_6-\tfrac{1}{3}\vece_7+\tfrac{1}{3}\vece_8.
\]

\newpage
\section{Special functions}

The Jacobi theta functions are defined as
\[
\varth_1(z,\tau)\Eqn{:=}
 i\sum_{n\in\bbZ}(-1)^n y^{n-1/2}q^{(n-1/2)^2/2},\nn\\
\varth_2(z,\tau)\Eqn{:=}
  \sum_{n\in\bbZ}y^{n-1/2}q^{(n-1/2)^2/2},\nn\\
\varth_3(z,\tau)\Eqn{:=}
  \sum_{n\in\bbZ}y^n q^{n^2/2},\nn\\
\varth_4(z,\tau)\Eqn{:=}
  \sum_{n\in\bbZ}(-1)^n y^n q^{n^2/2},
\]
where
\[
y=e^{2\pi iz},\qquad q=e^{2\pi i\tau}.
\]
We often use the following abbreviated notation
\[
\varth_k(\tau):=\varth_k(0,\tau).
\]
The Dedekind eta function is defined as
\[
\eta(\tau):=q^{1/24}\prod_{n=1}^\infty(1-q^n).
\]
The Eisenstein series are given by
\[
E_{2n}(\tau)
 =1-\frac{4n}{B_{2n}}\sum_{k=1}^{\infty}\frac{k^{2n-1}q^k}{1-q^k}
\]
for $n\in\bbZ_{>0}$. The Bernoulli numbers $B_k$ are defined by
\[
\frac{x}{e^x-1}\Eqn{=}\sum_{k=0}^\infty\frac{B_k}{k!}x^k.
\]
We often abbreviate $\eta(\tau),\,E_{2n}(\tau)$ as
$\eta,\,E_{2n}$ respectively.

Modular properties of the above functions are as follows:
\begin{align}
\varth_1(z,\tau+1)&=e^{\frac{\pi i}{4}}\varth_1(z,\tau),&
\varth_1(\tfrac{z}{\tau},-\tfrac{1}{\tau})
 &=e^{-\frac{3\pi i}{4}}\tau^{\frac{1}{2}}
  e^{\frac{\pi i}{\tau}z^2}\varth_1(z,\tau),\nn\\
\varth_2(z,\tau+1)&=e^{\frac{\pi i}{4}}\varth_2(z,\tau),&
\varth_2(\tfrac{z}{\tau},-\tfrac{1}{\tau})
 &=e^{-\frac{\pi i}{4}}\tau^{\frac{1}{2}}
  e^{\frac{\pi i}{\tau}z^2}\varth_4(z,\tau),\nn\\
\varth_3(z,\tau+1)&=\varth_4(z,\tau),&
\varth_3(\tfrac{z}{\tau},-\tfrac{1}{\tau})
 &=e^{-\frac{\pi i}{4}}\tau^{\frac{1}{2}}
  e^{\frac{\pi i}{\tau}z^2}\varth_3(z,\tau),\nn\\
\varth_4(z,\tau+1)&=\varth_3(z,\tau),&
\varth_4(\tfrac{z}{\tau},-\tfrac{1}{\tau})
 &=e^{-\frac{\pi i}{4}}\tau^{\frac{1}{2}}
  e^{\frac{\pi i}{\tau}z^2}\varth_2(z,\tau),\nn\\
\eta(\tau+1)&=e^{\frac{\pi i}{12}}\eta(\tau),&
\eta(-\tfrac{1}{\tau})
 &=e^{-\frac{\pi i}{4}}\tau^{\frac{1}{2}}\eta(\tau),\nn\\
E_{2n}(\tau+1)&=E_{2n}(\tau),&
E_{2n}(-\tfrac{1}{\tau})
 &=\tau^{2n}E_{2n}(\tau)\quad(n\ge 2).
\end{align}
%


\renewcommand{\section}{\subsection}
\renewcommand{\refname}

\end{document}